\documentclass{svjour2}                    
\usepackage{graphicx}
\usepackage{amssymb}
\usepackage{amsmath}

\def\dbar{{\mathchar'26\mkern-11mu{\rm d}}}

\journalname{}

\begin{document}

\title{Status of the Clausius inequality in classical thermodynamics}
\titlerunning{Clausius inequality}

\author{A.~V.~Gavrilov}

\institute{A.~V.~Gavrilov\at
              Department of Physics, Novosibirsk State
University, 2 Pirogov Street, Novosibirsk, 630090, Russia. \\
              \email{gavrilov19@gmail.com}           
}

\date{Received: date / Accepted: date}

\maketitle

\begin{abstract}

We present an analysis of the foundations of the well known Clausius
inequality. It is shown that, strictly speaking, the inequality is
not a logical consequence of the Kelvin-Planck formulation of the
second law of thermodynamics. Some thought experiments demonstrating
the violation of the Clausius inequality are considered. Also, a
reformulation of the Landauer's principle in terms of the Clausius
inequality is proposed. This version of the inequality may be
considered a consequence of the fluctuation theorem.

\end{abstract}

\section*{I. INTRODUCTION}

 This is a paper in classical thermodynamics. So, it is
uncommon: classical (or macroscopic) thermodynamics, unlike
statistical physics, is rarely a subject of a research paper today.
Taking into account the treatment of the Clausius inequality, it
must be more than uncommon. To avoid misunderstanding, the author
has to say that to refute the inequality (or to prove it) is not a
goal of this paper. The goal is to disclose a gap in the foundations
of classical thermodynamics and to exploit it.

The paper consists of two parts. In the first one (Sec. II), we show
that a hypothetical violation of the Clausius inequality does not
contradict the basic principles of thermodynamics, like the second
law. In other words, to test the Clausius inequality and to try to
invent a perpetuum mobile of the second kind is not the same. The
second part (Secs. III, IV) is devoted to another problems related
to the Landauer's principle. Despite the appearance, these two parts
are closely related. One of the reasons is that a hypothetical
violation of the Clausius inequality is considered in both, although
in different context. The summary is given below.

\subsection*{A. The Clausius inequality in macroscopic thermodynamics}

We show in II A that the Clausius inequality is based on an implicit
assumption about a process, which does not follow from any of the
known principles. We call this assumption \emph{ environment
independence}. Environment dependence is not exactly a new concept
in thermodynamics. The general idea is simple: while a system
undergoes a cycle, the environment is changed, the change should be
taken into account to get the thermodynamics right. Bennett in
\cite{B} used essentially the same argument to resolve the Maxwell's
demon paradox, although in this case it is convenient to put it in
terms of information.

If we do not assume environment independence, then a violation of
the Clausius inequality does not contradict the laws of
thermodynamics. It can be explained by \emph{adiabatic entropy
transfer} between thermodynamic systems (II C,D). A proper
definition of Clausius entropy in this case is discussed in II D.
Thermodynamics of an imaginary system capable of violating the
Clausius inequality is considered in II B and II E. Whether it can
be violated in a real process (by a macroscopic system) is a
difficult question which is not addressed here.

\subsection*{B. The Clausius inequality in information thermodynamics}

We call information thermodynamics the area of research related to
the Landauer's principle, Maxwell's demon, and similar topics. It is
almost obvious that the Clausius inequality \emph{can} be violated
in this context. Still, this fact does not attract much attention.
The reason is, the only methods used in information thermodynamics
today are the methods of statistical physics. In statistical
physics, the inequality does not play a significant role, and it is
possible to ignore its violation.

On the other hand, the Clausius inequality plays a fundamental role
in classical thermodynamics, as a basis of Clausius entropy. The
methods of classical thermodynamics can certainly be applied to
Maxwell's demon, to erasure of information etc., although not in a
straightforward way. However, the decade old paper of Ishioka and
Fuchikami \cite{IF} is the only attempt known to the author. So, he
decided to make the second attempt.

The proposed theory is \emph{nonstandard thermodynamics}, which is
an extension of familiar (standard) classical thermodynamics. The
formalism is explained in III B and the principles are discussed in
Sec. IV. In terms of nonstandard thermodynamics, the Landauer's
principle has a natural interpretation as a generalization of the
Clausius inequality (III C). This general inequality is proved in
III D by means of Hamiltonian dynamics, as a consequence of the
Crooks fluctuation theorem. (This is the only part of the paper
where methods of statistical physics are employed. A more
``classical'' proof is sketched in IV C.) The theory is applied to
erasure of information (III E), to the magnetization reversal (III
F), and to the Szilard engine (IV B,E).

There are some new results: a formula for the area of the hysteresis
loop (18), another formula for the dispersion of work (23), as well
as a specific version of the Clausius inequality (17). (A similar
but different inequality was found by Sagawa and Ueda
\cite[Eq.(3)]{SU}.) But the main goal was not to obtain new results,
it was to bridge the gap between modern statistical mechanics and
oldfashioned classical thermodynamics. So, we mostly use the
thermodynamic methods instead of traditional methods of statistical
physics. Technically, it means quite a different logic, but from a
more pragmatic viewpoint the main difference from the conventional
approach is in the notion of a process (discussed in III B).

\section*{II. THE STATUS OF THE CLAUSIUS INEQUALITY}
\subsection*{A. The Clausius inequality and the second law}

The Clausius inequality is a well known statement of classical
thermodynamics. Consider a system undergoing a cyclic process in
contact with a heat bath or with a sequence of baths, one at a time.
The Clausius inequality gives an upper bound for integrated reduced
heat

\begin{equation}
\oint\frac{\dbar Q}{T}\le 0.\label{cla}
\end{equation}

Here $\dbar Q$ is heat taken by the system from a bath at absolute
temperature $T$. (Note that in general the system itself is not
supposed to have a well defined temperature.) The inequality was
named after Rudolf Clausius, who introduced it in the famous 1865
paper \cite{Cla}.

One can prove the inequality either by means of classical
thermodynamics or statistical physics. But in this section we
consider thermodynamic methods only. We follow the argument of
Clausius, which may be found in many textbooks \cite{Fer}\cite{Fey}.
(There is an alternative argument made by Caratheodory \cite{Car}.
It is mentioned in II D, but in general the Caratheodory's method is
too narrow in scope for our purpose. We do not consider quasistatic
processes only.)

It is enough to consider a system undergoing a cyclic process in
contact with a single heat bath. In this case, the Clausius
inequality is simply $Q\le 0$, where $Q$ is heat taken from the bath
in a cycle. By the first law, $W=-Q$, where $W$ is work done on the
system. If $Q>0$, then (positive) heat is taken from a single bath
and converted to work in a cycle. It looks like a contradiction to
the second law, but actually it is not yet.

The Kelvin-Planck formulation of the second law states that \emph{
it is not possible to take heat from a single heat bath and convert
it to work in a cyclic process}. But a ``cyclic process'' in this
statement is not a cyclic process we have in mind. It is a process
whose \emph{only} net result is to take heat from a bath and convert
it to work \cite{Fer}\cite{Fey}. That is, not just one system, but
\emph{each} system (except for the bath) is supposed to undergo a
cycle. Usually, competent authors of  textbooks point this out.

This ambiguity in the meaning of the term ``cycle'' is a potent
source of confusion. To avoid it, we use the term \emph{global
process} when referring to all the thermodynamic systems, excluding
the heat baths. A \emph{process} or  \emph{local process} is always
related to a single system. (Which may be a bath.) In this
terminology, the Clausius inequality is about local cycles while the
second law is about global ones. It makes a difference.

To complete the argument, we have to assume that \emph{it is
possible to make the system undergo the process in such a way that
all the environment, with the exception of a heat bath, remains
unchanged}. Call a process \emph{environment independent} if this
assumption is true and \emph{environment dependent} otherwise.
 When a system $S$ undergoes an environment dependent
cycle, some other system $S^{\prime}$ undergoes a ``parallel''
process, which may be cyclic or not (Fig. 1 (a,b)). To make a global
cycle we have to bring $S^{\prime}$ to the original state. This may
require some work to be done on $S^{\prime}$ and some heat
$-Q^\prime$ to be taken from it (Fig. 1 (c)). What follows from the
second law in this case is not the inequality $Q\le 0$, it is the
inequality $Q+Q^\prime\le 0$.

\begin{figure}
  \centering
  \includegraphics{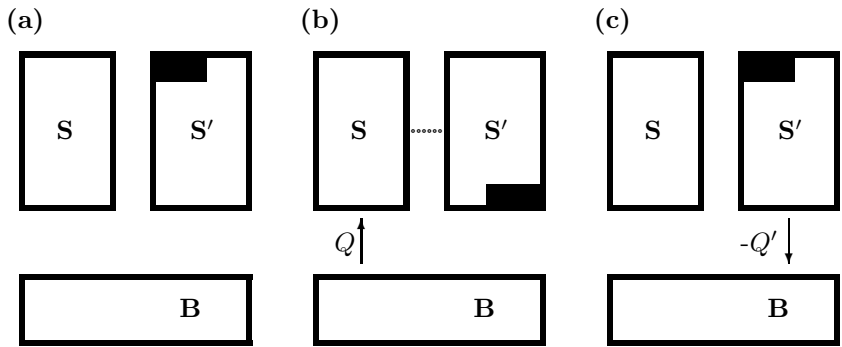}\\
  \caption{Environment dependent process. { \bf (a,b)} The system $S$ undergoes a cycle,
  taking heat $Q$ from the heat bath $B$. Simultaneously, the system
 $S^{\prime}$ undergoes a change due to interaction with $S$. {\bf (c)} $S^{\prime}$
 returns to the initial state, giving heat  $-Q^\prime$ to the bath.}
\end{figure}

Certainly, environment dependence implies  a sort of interaction
between the system and the environment. (Which is shown on Fig.1 (b)
schematically by a dotted line.) From thermodynamic viewpoint, it is
a peculiar interaction, for it does not involve heat or work.
However, it is not excluded \emph{a priory} by any of the known
principles. If every process is supposed to be environment
independent, we call it \emph{standard thermodynamics}. Basically,
it is thermodynamics  familiar from textbooks. If this assumption is
dropped, then what is left is \emph{weak thermodynamics}, which is
the subject of the rest of this section.

To make the argument down to earth, we have to consider at least one
particular environment dependent process. A system considered below
is the \emph{xenium engine}, which is an imaginary device capable of
violating the Clausius inequality. It is an interesting question to
what extent it is realistic and whether a similar device can exist
in nature, but this question is not relevant to the matter. The
engine obeys the second law but violates the Clausius inequality.
Whether it is realistic or not, this is enough to conclude that the
latter is not a consequence of the former.

\subsection*{B. The xenium engine}

The xenium engine is a kind of heat engine, with an imaginary gas
\emph{xenium} as the working body. For the sake of convenience it is
denoted by a ``chemical'' symbol ${\rm Xe }$. Xenium is an ideal
gas. A molecule of xenium can be in one of two states, denoted by
${\rm Xe_a}$ and ${\rm Xe_b}$. (With the same energy levels.) A
single molecule can never change its state. However, two
sufficiently close molecules may exchange their states

\begin{equation}
{\rm Xe_a}+{\rm Xe_b^{\prime}} \longleftrightarrow {\rm Xe_b}+{\rm
Xe_a^{\prime}}.\label{xexe}
\end{equation}

(Here ${\rm Xe}$ is a molecule and ${\rm Xe^{\prime}}$ is another
one.) The total number $N_a(N_b)$ of the ${\rm Xe_a}({\rm Xe_b})$
molecules remains constant, so xenium is a mixture of two gases to
some extent.

The xenium engine is a cylinder with a piston which moves without
friction (Fig. 2). The wall opposite to the piston is adiabatic. It
is also thin (in a sense explained below). The cylinder is divided
in two by a semipermeable partition. The ${\rm Xe_a}$ molecules can
penetrate it while the ${\rm Xe_b}$ ones can not. The space between
the thin wall and the partition, called a \emph{camera}, is filled
with a mixture of ${\rm Xe_a}$ and ${\rm Xe_b}$. Another part of the
cylinder is filled with pure ${\rm Xe_a}$.
\begin{figure}
  \centering
  \includegraphics{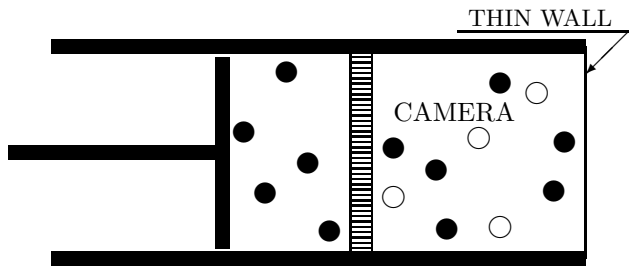}\\
  \caption{The xenium engine}
\end{figure}

Consider first a single engine in contact with a heat bath at
temperature $T$. The piston moves in a quasistatic (hence
reversible) process. The pressure $P$ on the piston is then equal to
the partial pressure of ${\rm Xe_a}$ in the camera. By the
Gay-Lussac law,
$$P=N_ak_BT/V,$$
where $k_B$ is the Boltzmann constant and $V$ is the volume between
the piston and the thin wall. It is convenient to introduce the
dimensionless volume $v=V/V_0\ge 1$, where $V_0$ is the (constant)
volume of the camera. Work in the process is
$$\dbar W=-PdV=-N_ak_BTd\ln v.$$
The internal energy of an ideal gas does not depend on volume, hence
$\dbar Q=-\dbar W$ and

\begin{equation}
\frac{\dbar Q}{T}=N_ak_Bd\ln v.\label{QT}
\end{equation}

Consider now two xenium engines connected as on Fig. 3. Each engine
is in contact with a particular heat bath. The adiabatic wall
separating the engines is so thin  that xenium molecules in one
camera may interact with molecules in another camera. The
interaction looks somewhat similar to  diffusion (Fig. 4). However,
there is no real flow of molecules through the wall. Molecules
simply change their states, so the numbers $N_a$ and $N_b$ aren't
actually conserved. (This will be transparent if we assume that
xenium on the two sides of the wall are two different isotopes.)

\begin{figure}
  \centering
  \includegraphics{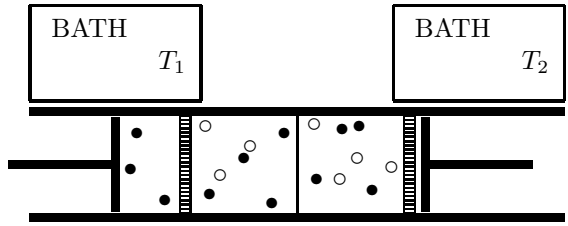}\\
  \caption{Pair of xenium engines}
\end{figure}

\begin{figure}
  \centering
  \includegraphics{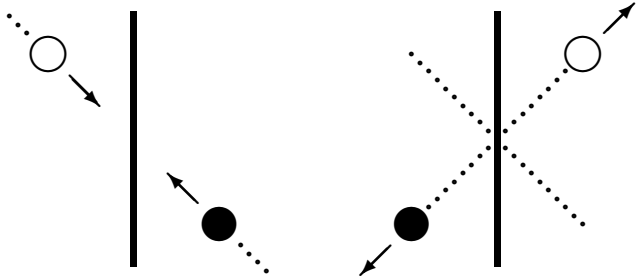}\\
  \caption{Interaction through the thin wall. A molecule of ${\rm Xe_a}$ (black ball)
  turns to ${\rm Xe_b}$ (white ball), and vice versa.}
\end{figure}

To distinguish variables related to different engines we use
subscripts `1' and `2'. While the total number of ${\rm Xe_a}$
molecules $N_{a,1}+N_{a,2}$ remains a constant, the summands became
functions of the variables $v_1$ and $v_2$. It is not difficult to
find this functions explicitly. Denote by $z$ the quotient of the
concentrations in the camera:
$$z=[{\rm Xe_a}]\slash [{\rm Xe_b}]=N_a\slash vN_b.$$
Then $N_a=N\slash(1+v^{-1}z^{-1}),$ where $N=N_a+N_b$ is a constant
for each engine.

By symmetry, the equilibrium constant of the reaction \eqref{xexe}
is one, hence it comes to equilibrium when $z_1=z_2$. In a
quasistatic process this equality must hold all the time. In the
case $N_1=N_2=N_{a,1}+N_{a,2}=N$ we have
$$z_1=z_2=\frac{1}{\sqrt{v_1v_2}}
,\,N_{a,1}=\frac{N\sqrt{v_1}}{\sqrt{v_1}+\sqrt{v_2}},\, \frac{\dbar
Q_1}{T_1}= \frac{2k_BN}{\sqrt{v_1}+\sqrt{v_2}}d\sqrt{v_1}.$$

Clearly, $\dbar Q_1/T_1$ is \emph{not} an exact differential. On the
other hand, the sum
$$\frac{\dbar Q_1}{T_1}+\frac{\dbar Q_2}{T_2}=2k_BNd\ln(\sqrt{v_1}+\sqrt{v_2})$$
is an exact differential. (Which is also true for any choice of
parameters.)

The conclusions are as follows. The xenium engine undergoes a
reversible process, in which the reduced heat is not an exact
differential. Thus, the Clausius inequality cannot be true for all
the cycles. On the other hand, the sum $\dbar Q_1\slash T_1+\dbar
Q_2\slash T_2$ is an exact differential, which means that no
contradiction to the second law is possible. So, in this situation
the Clausius inequality is not really a consequence of the second
law.

\subsection*{C. The weak Clausius inequality}

In weak thermodynamics, the classical Clausius's argument has to be
modified.  First of all, we have to consider a global cycle, not a
local one. There can be many systems, interacting with each other in
any possible way. Each of these systems is supposed to undergo a
cyclic process, except for heat baths. Let us introduce one more
heat bath at temperature $T_0$.  We can then restore every bath,
save this one, to its initial state by giving it heat at the expense
of heat taken from the exceptional bath. To make this process
reversible we use Carnot engines ($=$ environment independent
reversible cyclic devices). Now, as the ``intermediate'' baths do
not take essential part in the process, we can ignore them (Fig. 5).

\begin{figure}
  \centering
  \includegraphics{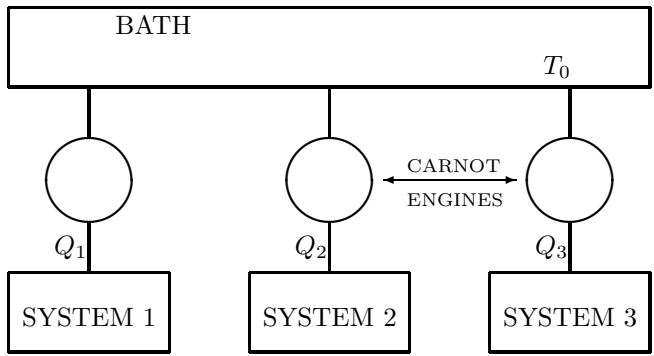}\\
  \caption{Three systems connected to a heat bath.}
\end{figure}

By a simple calculation,

$$\sum_i \oint\frac{\dbar Q_i}{T_i}=\frac{Q_0}{T_0},$$

where the sum is taken over all the systems and $Q_0$ is heat taken
from the exceptional bath. By the second law, the right hand side
cannot be positive. Thus, we have the \emph{weak Clausius
inequality}

\begin{equation}
\sum_i \oint\frac{\dbar Q_i}{T_i}\le 0.\label{cla1}
\end{equation}

Call a global cyclic process \emph{reversible} if \eqref{cla1} is an
exact equality or the difference between the left hand side and zero
is negligible. A local process, cyclic or not, is reversible if it
may be a part of a reversible global cycle. (This is a broad
definition of reversibility, which does not imply equilibrium. For
example, a diamond at standard temperature and pressure undergoes a
reversible process, despite being far from thermodynamic
equilibrium.)

Following Clausius, we can define the \emph{total} entropy by the
equality

\begin{equation}
dS=\sum_i \frac{\dbar Q_i}{T_i}, \label{S1}
\end{equation}

in a reversible global process. (This is a consistent definition
because of \eqref{cla1}. This total entropy is the entropy of
``genuine'' systems; entropy of heat baths is not included. For this
reason, it is not a constant even in a reversible process.) The
definition of the entropy of an individual system is considered
below.

\subsection*{D. The Clausius entropy}

In classical thermodynamics, the very definition of entropy is based
on the Clausius inequality. If the inequality is not valid, this
definition has to be modified. Fortunately, the modification is
rather obvious. All we have to do is to make explicit an implicit
environment independence assumption. Thus, the Clausius entropy $S$
of a system is defined by $dS=\dbar Q\slash T$ in a \emph{reversible
environment independent} process.

In general, we have $dS\neq \dbar Q\slash T$, even if a process is
reversible. Call the difference $\dbar{\mathfrak S}$ \emph{entropy,
transferred adiabatically} to the system. So,

\begin{equation}
dS=\frac{\dbar Q}{T}+\dbar{\mathfrak S}.\label{S2}
\end{equation}

The total entropy defined by \eqref{S1} must be equal to the sum
$\sum_i S_i$ of ``local'' entropies. From \eqref{S2}, we have
$$\sum_i \dbar{\mathfrak S}_i=0.$$
The latter equality is a justification of the term ``transfer''.
Entropy does not come from nowhere, it comes from the environment.
If $T={\rm const}$, then
$$\Delta F=W-T{\mathfrak S},$$
where $F=U-TS$ is the Helmholtz free energy. Under close
examination, the term $-T{\mathfrak S}$ on the right hand side is
\emph{chemical work}. (Note that it is done on a \emph{closed}
system, which is somewhat uncommon.)

For a general (not reversible) process the equality \eqref{S2} turns
into an inequality

\begin{equation}
dS\ge \frac{\dbar Q}{T}+\dbar{\mathfrak S}. \label{S3}
\end{equation}

Of course, in this case $\dbar{\mathfrak S}$ cannot be defined as a
difference between $dS$ and $\dbar Q\slash T$. The definition is
$\dbar{\mathfrak S}=-\dbar{\mathfrak S}_{env}$, where
$\dbar{\mathfrak S}_{env}$ is entropy transferred to the environment
(which may be supposed to undergo a reversible process).
 Taking the integral of the both sides of \eqref{S3}  over a cycle, we have the
following generalization of the Clausius inequality

\begin{equation}
\oint\frac{\dbar Q}{T}\le-{\mathfrak S}.\label{cla2}
\end{equation}

The standard Clausius inequality \eqref{cla} can be violated if
${\mathfrak S}<0$, i.e. if entropy is transferred to the environment
adiabatically.

So far, we followed Clausius. It is also instructive to take another
view, more close to Caratheodory's. The state of a system can be
described by the entropy $S$ and the parameters $x_1,\dots,x_n$. Let
$U(S,x_1,\dots,x_n)$ be the internal energy as a function of these
parameters. In an adiabatic  process, $dU=\dbar W$. Caratheodory has
assumed that work (in a reversible process) is $\dbar W=\sum_i
p_idx_i,$ for some functions $p_i$. It follows that

\begin{equation}
TdS=\sum_i\left(p_i-\frac{\partial U}{\partial x_i}\right)dx_i.
\label{carat}
\end{equation}

The point is, in an environment dependent adiabatic process, $dS\neq
0$, i.e. entropy is not conserved. In fact, \emph{nothing} is
conserved: any two states may be connected by an adiabatic process,
as soon as environment dependent effects are present. This
invalidates the central postulate of Caratheodory's theory. To make
sense of entropy, we have to consider environment independent
processes separately; in this respect, there is no much difference
with Clausius's approach. In an environment independent adiabatic
process $dS=0$, hence  the sum on the right hand side of
\eqref{carat} is zero. However, it does \emph{not} follow that
$p_i=\partial U\slash\partial x_i$, because in this case the
differentials $dx_i$ are not linearly independent. For this reason,
in a general process this sum (which is chemical work, with the
opposite sign) is not zero.

In this argument, the difference between a \emph{reversible} process
(as it is defined in II C) and an \emph{equilibrium} process is very
important.  Almost by definition, an equilibrium state may only
depend on those parameters which can be changed by an external agent
at will, such as volume, magnetic induction etc. It means that any
equilibrium process is environment independent (and there can be no
linear dependence between the differentials of the parameters). So,
no adiabatic entropy transfer is possible in an equilibrium process.

\subsection*{E. The entropy of the xenium engine}

The Clausius entropy of the xenium engine is more tricky to find
than the Boltzmann-Gibbs entropy. (We have less information because
of the black box method inherent in classical thermodynamics.) Of
course, it is the same entropy in the end. We do the calculation
below, in order to show the tools at work, and to point out some
peculiarities. For the sake of simplicity, we consider isothermal
processes only.

The system has two parameters. Convenient parameters are the
dimensionless volume $v$, and the number of ${\rm Xe_a}$ molecules
$N_a$. We have a problem with the second one: in any environment
independent process $N_a$ is a constant. So, we cannot measure the
entropy difference between states with different $N_a$ values.

However, it is not a fault of the method, it is a fault of the
model.  Let us suppose that there  exists a catalyst which makes
xenium molecules undergo spontaneous transitions ${\rm
Xe_a}\leftrightarrow{\rm Xe_b}$. We can add the catalyst to or
remove it from the camera as necessary. With catalyst in the camera,
we have $[{\rm Xe_a}]=[{\rm Xe_b}]$, which means $N_a=N/(1+v^{-1})$.
Otherwise, $N_a$ is a constant. (To reach states with $N_a>N\slash
2$, we need a still more elaborated model, which is not considered
here.)

In any quasistatic process, the equality \eqref{QT} is true
regardless of environment dependence. Thus,
$$dS=N_ak_Bd\ln v,$$

in an environment independent process. Environment independence in
this case implies either $N_a={\rm const}$ or $N_a=N/(1+v^{-1})$.
The right hand side is not an exact differential, but this does not
matter. (Because any environment independent cycle is trivial: it
does not enclose positive area on the $v$ vs. $N_a$ plane.) Taking
integral over the allowed processes, we have
$$S=N_ak_B\ln(v\slash N_a)-N_bk_B\ln N_b+C,$$
where $N_b=N-N_a$ and $C$ is a constant. This entropy coincides with
the Boltzmann-Gibbs entropy
$$S=-N_ak_B\ln[{\rm Xe_a}]-N_bk_B\ln[{\rm Xe_b}],$$
as expected.

For an environment dependent process, $dS\neq\dbar Q\slash
T=N_ak_Bd\ln v$. The difference is
$$\dbar{\mathfrak S}=dS-\dbar Q\slash T=-k_B\ln z dN_a,$$
where $z=N_a\slash vN_b$.

Again, it was expected. The chemical work in the process is
$$-T\dbar{\mathfrak S}=\mu_adN_a+\mu_bdN_b=(\mu_a-\mu_b)dN_a,$$
where $\mu_a$ and $\mu_b$ are the chemical potentials of ${\rm
Xe_a}$ and ${\rm Xe_b}$ respectively. For an ideal gas
$$\mu_a-\mu_b=k_BT\ln[{\rm Xe_a}]-k_BT\ln[{\rm Xe_b}]=k_BT\ln z.$$

The author would like to make a remark. Once we introduce a
catalyst, it become obvious that the xenium engine works out of
equilibrium. Xenium, as a thermodynamic system, can only be in
equilibrium if $[{\rm Xe_a}]=[{\rm Xe_b}]$. Otherwise, we can add a
catalyst and observe an irreversible reaction. As it was noted at
the end of II D, we could not expect a violation of the Clausius
inequality in an equilibrium process.

\section*{III. THE LANDAUER'S PRINCIPLE}

\subsection*{A. The original Landauer's principle}

The Landauer's principle was proposed by Rolf Landauer in
\cite{Lan}. Apparently, this work did not attract much attention
before 1982, when Bennet used the principle to explain the
paradoxical behavior of the Szilard engine \cite{B}. By now, there
exists quite an extensive literature on this subject. Some of more
recent or more relevant works are
\cite{Berut}\cite{Dill}\cite{Pie}\cite{Mar}\cite{SU}.

According to the principle, \emph{erasure of information is
accompanied by heat generation} \cite{Lan}. The quantitative
formulation may be presented as an inequality
$$Q_{dis}\ge k_BTI,$$
where $Q_{dis}$ is heat dissipated in the process and $I$ is the
erased information (in nats). Of course, to make it a precise
statement one has to specify the definitions of $Q_{dis}$ and $I$,
but we avoid this technicalities. It is enough to say that $I$ is
usually defined as Shannon entropy. (It may also be von Neumann
entropy, but we only consider classical information here.)

The author is convinced that it is much more natural to interpret
the Landauer's principle as a statement in classical thermodynamics
than a statement in statistical physics. But in view of
thermodynamics, the original formulation is really weird. Classical
thermodynamics does not deal with ensembles, so Shannon entropy does
not make much sense. To give a reasonable thermodynamic formulation
of the principle we need another measure of information. The
proposed measure is \emph{thermodynamic information}, introduced
below.

\subsection*{B. Nonstandard thermodynamics}

Apparently, classical thermodynamics is considered by many a kind of
adaptation of statistical physics for engineers. The author does not
share this opinion. Classical thermodynamics is a valuable part of
theoretical physics, which proved to be useful on many occasions. It
may be useful in information thermodynamics, too. To follow this
route we have to adopt the black box method and to employ Clausius
entropy instead of Boltzmann-Gibbs  entropy.

However, there is an obstacle. One of the cornerstones of classical
thermodynamics is determinism. Stochastic phenomena of any kind are
treated as fluctuations, which are beyond the scope of
thermodynamics itself. But many of the processes usually considered
in information thermodynamics (such as the Szilard engine cycle) are
not deterministic. So, we have to \emph{extend} classical
thermodynamics.

The proposed extension is \emph{nonstandard thermodynamics}. While
the traditional, or \emph{standard} classical thermodynamics deals
with deterministic processes only, in nonstandard thermodynamics we
may consider a process which can be observed with some probability.
The fundamental characteristics of a process are work $W$ done by an
external agent, heat $Q$ taken from a heat bath(s), and the
\emph{probability of the process} $P$. What is important, this
probability can be measured, at least in principle, by trying the
process many times. So, the probability of a process is a property
of a system \emph{as a black box}: we do not have to know anything
about the internal physics to measure it. In this respect, it isn't
different from heat or work.

When we speak about a probability, we imply a number of
alternatives. It should be noted that for a particular process the
alternatives may be different, depending on the circumstances. (This
is usually referred to as feedback control.) But the probability
itself is supposed to be the same, so we may consider it a
characteristic of a process.

The author would like to point out that this approach to a process
is radically different from what one may find in the literature. The
tradition is to consider all the possibilities at once
\cite{B}\cite{Fah}\cite{Pie}\cite{Par}\cite{MNV}, so ``work (heat)''
actually means ``the mean value of work (heat)'', averaged over all
the possible alternative processes. This is natural in the context
of statistical physics, where the main tool is an ensemble. However,
it would not be so natural in the context of classical
thermodynamics. For this reason, in nonstandard thermodynamics
\emph{we do not consider ensembles}. We consider a single system
undergoing a single process a single time. Moreover, the second time
the outcome of the process is supposed to be exactly the same (with
some exceptions discussed in IV A). In this respect, nonstandard
thermodynamics is not very much different from the standard one.

For example, the Szilard engine, considered in IV B, can undergo one
of two different cyclic processes, depending on where the working
particle gets trapped. In the both cases, the probability is
$P=1\slash 2$ and work is $W=-k_B T\ln 2$. But for a ``skewed''
version of the engine the works (and probabilities) are different,
so it is more convenient to consider these two processes separately.
Note that as far as we are interested in a single process, it makes
no difference if there is any feedback control or not.
(Occasionally, we consider a family of processes ${\mathcal A}_i$,
which may be observed in a certain situation with probabilities
$P_i$, such that $\sum_i P_i=1$. We call it a \emph{mixed process},
denoted by  $\sum_i P_i{\mathcal A}_i$.)

We assume that for each process ${\mathcal A}$ there exists an
\emph{opposite} process ${\mathcal A}^\dagger$, which consists of
the same states taken in the reverse order. The \emph{backward
probability} $P^\dagger$ of a process is the probability of the
opposite process, $P^\dagger({\mathcal A})=P({\mathcal A}^\dagger)$.
A process is called \emph{balanced} if $P\ge P^\dagger$,
\emph{deterministic} if $P=1$ and \emph{bideterministic} if
$P=P^\dagger=1$.

We assume also that heat taken by a system in a ``trivial'' cycle
like ${\mathcal A}^\dagger{\mathcal A}$ cannot be positive. One can
see that it is equivalent to the inequality $\dbar Q+\dbar
Q^\dagger\le 0$, where (infinitesimal) heat $\dbar Q$ is taken in a
part of the process and $\dbar Q^\dagger$ is taken in the
corresponding part of the opposite process. Call a process
${\mathcal A}$ \emph{reversible} if no heat is taken in the cycle
${\mathcal A}^\dagger{\mathcal A}$. In this case, the opposite
process a \emph{reverse} process, ${\mathcal A}^{-1}={\mathcal
A}^\dagger$.

The author would like to point out that this definition is
\emph{different} from the definition of a reversible process in
standard thermodynamics. (Formally, it is Carnot reversibility in
terms of \cite[Appendix C]{Jay}, taken in nonstandard context.) For
example, a process with the probabilities $P=1,P^\dagger<1$ is
irreversible in standard thermodynamics whether it is reversible in
the above sense or not. (Parrondo \cite{Par} used the term
\emph{quasiirreversible}.) A  process which is reversible in
standard thermodynamics is a bideterministic reversible process.

If ${\mathcal A}{\mathcal B}$ is a composition of two processes
(${\mathcal A}$ followed by ${\mathcal B}$), then $P({\mathcal
A}{\mathcal B})=P({\mathcal A})P({\mathcal B})$. It follows that the
function of a process $I$, defined by
$$I=\ln\frac{P^\dagger}{P},$$
is additive, i.e. $I_{{\mathcal A}{\mathcal B}}=I_{\mathcal
A}+I_{\mathcal B}$. We will often write it in the integral form,
$I=\int \dbar I$. Call it \emph{thermodynamic information} created
in the process (measured in nats). Thermodynamic information is
\emph{not} information in the sense of Shannon. We will see,
however, that there is a good reason to consider it a thermodynamic
analog of Shannon information.

At this point, the author would like to make a few remarks.
 To avoid possible confusion, in  Secs. III, IV we do not
take into account any environment dependence effects discussed in
Sec. II. It simply means  that in standard thermodynamics the
Clausius inequality \eqref{cla} is valid. However, it is \emph{not}
valid in nonstandard thermodynamics (for reasons explained below).
Moreover, it is often convenient to interpret a violation as a
consequence of a (fictitious) adiabatic entropy transfer.

The author did his best to find an appropriate language for the
classical counterpart of information thermodynamics, but the two
main contributions were made be the others. Clausius entropy was
introduced in information thermodynamics by Ishioka and Fuchikami
\cite{IF}. The relation between Clausius entropy and thermodynamic
information was, in essence, found by Kawai et.al. \cite{KPV}. So,
this work is not actually as original as it looks.

\subsection*{C. The Clausius inequality and the Landauer's principle}

In nonstandard thermodynamics, the Clausius inequality is different
from the familiar one \eqref{cla}. It is

\begin{equation}
\oint\frac{\dbar Q}{T}\le k_B I, \label{claI}
\end{equation}

where $I$ is thermodynamic information created in the cycle. The
inequality \eqref{claI} is both more general and more precise then
\eqref{cla}. First of all, \eqref{cla} is not true if $P<1$ (some
examples are considered below). It simply means that standard
thermodynamics cannot be applied to a process which is not
deterministic, which must not be a surprise. On the other hand, the
both inequalities can be applied to a deterministic cyclic process.
But in this case \eqref{claI} is stronger than \eqref{cla}, because
$I\le 0$.

If a cycle ${\mathcal A}$ is reversible (in the sense of III B),
then we have
$$k_BI_{{\mathcal A}}=-k_BI_{{\mathcal A}^{-1}}\le -
\oint_{{\mathcal A}^{-1}}\frac{\dbar Q}{T}=\oint_{{\mathcal
A}}\frac{\dbar Q}{T}\le k_BI_{{\mathcal A}}.$$

So, in this case the inequalities are in fact equalities. Repeating
the standard argument by Clausius, we have the following definition
of the Clausius entropy $S$

\begin{equation}
dS=\frac{\dbar Q}{T}-k_B\dbar I. \label{SI}
\end{equation}

In a general (not reversible) process, the equality turns into an
inequality

\begin{equation}
dS\ge\frac{\dbar Q}{T}-k_B\dbar I. \label{SI1}
\end{equation}

From \eqref{SI1} and the first law of thermodynamics $dU=\dbar
W+\dbar Q$, we have
$$dF\le \dbar W-SdT+k_BT\dbar I,$$

where $F=U-TS$ is the Helmholtz free energy. In the case of an
isothermal process it is equivalent to

\begin{equation}
W\ge \Delta F-k_B TI. \label{WI}
\end{equation}

(This inequality is known \cite[Eq. (8)]{KPV}.)

Consider a process which is isothermal and deterministic, but not
bideterministic. The dissipated heat,  as it is defined \emph{in
standard thermodynamics} is $Q_{dis}=W-\Delta F$. Thus,
$$Q_{dis}\ge k_BT|I|,$$

where $I=\ln P^\dagger<0$. We will see in III E  that in a standard
erasure of information process $|I|$ is, basically, erased
information in the sense of Landauer. This gives us a reason to
consider \eqref{claI} a \emph{reformulation} of the Landauer's
principle.

The original Clausius inequality \eqref{cla} can be derived from
classical thermodynamics or from statistical physics \cite{J}. The
same is true for the general inequality \eqref{claI}. In this case,
the author prefers statistical physics. (In IV C, the inequality is
derived from the second law in a specific formulation. However, this
formulation is by no means obvious or intuitive. So, this version of
the second law needs a convincing proof for itself.)

\subsection*{D. The Clausius inequality and the Crooks fluctuation theorem}

This is the only part of the paper devoted to statistical physics.
We prove a version of the Jarzynski equality (or the integral
fluctuation theorem), which is similar to the equality found
recently by Sagawa and Ueda \cite[Eq.(3)]{SU}. We use Hamiltonian
dynamics instead of stochastic dynamics, but it is not the main
difference. The result of Sagawa and Ueda is useless for our
purpose, because their definition of information is completely
different. (Apparently, this information cannot be defined properly
in the black box conditions.) The principal tool is the Crooks
fluctuation theorem, but, as was pointed out in \cite{SU}, basically
any version of the detailed fluctuation theorem can be employed.
(The relevant references can be found in \cite{SU}\cite{JJ}.)

A system in contact with a heat bath can be described by the
Hamiltonian
$$H(x,\lambda)+H_{I}(x,y)+H_{B}(y,\mu).$$
Here $x$ and $y$ are points in the phase space of the system and the
bath respectively, and $\lambda$ is a work parameter. In this sum,
$H$ is the energy of the system itself, $H_{I}$ is the interaction
energy and $H_{B}$ is the energy of the bath.

We do not want to restrict ourselves to isothermal processes. The
standard approach in this case is to introduce many baths, as in
\cite{J}\cite{SU}. But this is not very convenient for technical
reasons. Instead, we consider a single ``bath'' with variable
temperature. It is a system with very short relaxation time and
large specific heat. An external agent can change the temperature
$T$ of this system adiabatically, by varying the parameter $\mu$.

Consider a protocol $(\lambda(t),T(t)),$ where time $t$ is in the
interval $0\le t\le \tau$. In general, there are many processes with
the same protocol. A system which is not ergodic can be in different
(macroscopic) states at $t=0$. Moreover, the evolution of the system
may be indeterministic; in this case there can be many processes
with the same initial state.

A particular process ${\mathcal A}$ can be described by the initial
distribution $\rho_F(x)$ at $t=0$, and some set of trajectories. A
trajectory ${\mathcal X}=\{x(t):0\le t\le \tau\}$ is a curve in the
phase space. We write ${\mathcal X}\in{\mathcal A}$ if a system
which goes along this trajectory is considered to be undergoing the
process ${\mathcal A}$. The opposite process ${\mathcal A}^\dagger$
is described by the backward protocol $(\lambda(\tau-t),T(\tau-t))$,
another initial distribution  $\rho_B(x)$, and the conjugate set of
trajectories ${\mathcal A}^\dagger=\{{\mathcal X}^\dagger:{\mathcal
X}\in{\mathcal A}\}$. (Here ${\mathcal X}^\dagger=\{x(\tau-t)^*:0\le
t\le \tau\}$, where $x^*$ is obtained from $x$ by reversing the
momenta.)

Let $\mathcal{P}_F[{\mathcal X}]$  be the distribution of
trajectories (under the forward protocol), given the initial phase
space distribution $\rho_F(x)$, and $\mathcal{P}_B[{\mathcal X}]$ be
the distribution of trajectories under the backward protocol, given
the initial distribution $\rho_B(x)$. Then
$$P=P({\mathcal A})=\int_{{\mathcal A}}\mathcal{P}_F[{\mathcal X}],\,
P^\dagger=P({\mathcal A}^\dagger)=\int_{{\mathcal
A}^\dagger}\mathcal{P}_B[{\mathcal X}]=\int_{{\mathcal
A}}\mathcal{P}_B[{\mathcal X}^\dagger].$$

For a particular trajectory ${\mathcal X}$, denote by $x_0=x(0)$ the
beginning and by $x_1=x(\tau)$ the end.  Let
$\mathcal{P}_F[{\mathcal X}\mid x_0]$ and $\mathcal{P}_B[{\mathcal
X}\mid x_0]$ be the conditional distributions of trajectories, such
that
$$\mathcal{P}_F[{\mathcal X}]=\mathcal{P}_F[{\mathcal X}\mid x_0]\rho_F(x_0),\,
\mathcal{P}_B[{\mathcal X}]=\mathcal{P}_B[{\mathcal X}\mid
x_0]\rho_B(x_0).$$ The fraction
$$\mathcal{F}[{\mathcal X}]=\frac{\mathcal{P}_F[{\mathcal X}\mid x_0]}
{\mathcal{P}_B[{\mathcal X}^\dagger\mid x_1^*]}$$

plays an important role in statistical mechanics. It is
multiplicative in the following sense: if a trajectory ${\mathcal
X}={\mathcal X}_1{\mathcal X}_2$ consists of two parts ${\mathcal
X}_1$ and ${\mathcal X}_2$, then
 $\mathcal{F}[{\mathcal X}]=\mathcal{F}[{\mathcal
X}_1]\mathcal{F}[{\mathcal X}_2]$.

For an isothermal process, we have

\begin{equation}
\mathcal{F}[{\mathcal X}]=e^{-Q\slash k_BT},\, Q=\int_{{\mathcal
X}}\frac{\partial H}{\partial x}dx, \label{crooks}
\end{equation}

by the Crooks fluctuation theorem \cite[Eq. 9]{Cr}. (Here $Q$ is,
technically, stochastic heat, not thermodynamic heat. The difference
is explained in IV A.) A generalization to nonisothermal processes
is completely obvious due to multiplicativity

\begin{equation}
\mathcal{F}[{\mathcal X}]=\exp\left(-\frac{1}{k_B}\int_{{\mathcal
X}}\frac{\dbar Q}{T}\right). \label{Crooks}
\end{equation}

( Jarzynski used this equality in \cite[Eq.(6)]{J}, but he did not
write it down explicitly. Sagawa and Ueda used a similar equality
\cite[Eq.(11)]{SU}.)

Denote by
$$s=-k_B\ln(\rho(x)\slash\rho_0(x))$$
a \emph{stochastic entropy}, where $x$ is a point in the phase
space, $\rho$ is a distribution and $\rho_0$ is the Liouville
measure. (The measure is unique up to a factor, hence $s$ is well
defined up to a constant.) This ``entropy'' is a function of a
microstate and a macroscopic state together. For example, if the
distribution $\rho$ is canonical then $s=(H(x)-F)\slash T$, where
$F$ is the Helmholtz free energy.

The \emph{stochastic entropy production}
$$\sigma=\Delta s-\int_{{\mathcal X}}\frac{\dbar Q}{T}+k_BI$$
depends on the trajectory ${\mathcal X}$ as well as the initial and
final states. (Note that it is entropy production in nonstandard
thermodynamics. In standard thermodynamics, the term  $k_BI$ is not
included.)  Assuming $\rho_B(x^*)=\rho_B(x)$, we have $e^{-\Delta
s\slash k_B}=\rho_B(x_1^*)\slash\rho_F(x_0),$ hence by
\eqref{Crooks}
$$e^{-\sigma\slash k_B+I}=\frac{\rho_B(x_1^*)}{\rho_F(x_0)}\mathcal{F}[{\mathcal
X}]^{-1}=\frac{\mathcal{P}_B[{\mathcal X}^\dagger]}
{\mathcal{P}_F[{\mathcal X}]}.$$

Thus, the average of the exponent in the process is
$$\left\langle e^{-\sigma\slash k_B+I}\right\rangle_{\mathcal{A}}=
\frac{1}{P}\int_{\mathcal{A}}e^{-\sigma\slash
k_B+I}\mathcal{P}_F[{\mathcal
X}]=\frac{1}{P}\int_{\mathcal{A}}\mathcal{P}_B[{\mathcal
X}^\dagger]=\frac{P^\dagger}{P}.$$

We have the following version of the Jarzynski equality (compare to
\cite[Eq.(11)]{J} and \cite[Eq.(3)]{SU})

\begin{equation}
\left\langle e^{-\sigma\slash k_B}\right\rangle=1.\label{J}
\end{equation}

 If the process is isothermal and the distributions $\rho_F$, $\rho_B$
are canonical (but not necessary ergodic), then $\sigma=(W-\Delta
F)\slash T+k_BI$ and the equality become
$$\left\langle e^{-W\slash k_BT} \right\rangle=\frac{P^\dagger}{P}e^{-\Delta F\slash k_BT}.$$
In this form it was found by Kawai et.al. \cite[Eq.(7)]{KPV}.

The Clausius inequality \eqref{claI} is a simple consequence of
\eqref{J}. By the Jensen inequality, we have
$\langle-\sigma\rangle\le 0$. If the process is a cycle then
$\left\langle\Delta s\right\rangle=0$, hence

\begin{equation}
\left\langle \oint\frac{\dbar Q}{T}\right\rangle\le k_B I.
\label{claI2}
\end{equation}

In classical thermodynamics, the fluctuation are supposed to be
ignored, so we can replace the average by the actual value to obtain
\eqref{claI}. (A bit more detailed argument is in IV A.)

\subsection*{E. Erasure of information}

The applications of the inequality \eqref{claI} in this section are
two related but different processes: erasure of information and
magnetization reversal. Erasure of information is a standard test of
the Landauer's principle, so we can make a comparison between the
inequality and the principle to see what is common and what is
different.

The system under consideration is a 1-bit memory device, which is
normally in one of two states, called ``zero'' and ``one''. A
convenient model is a classical particle in a double-well potential
\cite{IF}\cite{Pie}\cite{Dill}, but actually the physics of the
device is not relevant to the analysis.  All we need is the
assumption that it obeys the inequality \eqref{claI}.

 Erasure of information is \emph{two} different processes. One of the
processes, denoted by ${\mathcal A}$, is a cycle which begins and
ends in the ``zero'' state. The other process ${\mathcal B}$ has the
initial state ``one'' and the final state ``zero''(Fig. 6 (a)). The
both processes are supposed to be isothermal and deterministic (that
is, information is erased for sure).

However, the opposite processes ${\mathcal A}^\dagger$ and
${\mathcal B}^\dagger$ are not deterministic. The argument may be as
follows. If an external agent ``knows'' which of the processes,
${\mathcal A}$ or ${\mathcal B}$, is going on, then this information
remains somewhere after a process is complete. In this case,
information would not be ``erased''. It means that the processes
must ``look'' the same for an external agent. The opposite processes
look the same as well, and have a common initial state. So, an
external agent does not have any means to make the system undergo,
say, the process ${\mathcal A}^\dagger$ for sure: the system has
free choice between ${\mathcal A}^\dagger$ and ${\mathcal
B}^\dagger$ (Fig. 6 (b)). Consequently, $P({\mathcal
A}^\dagger)+P({\mathcal B}^\dagger)=1$.

From \eqref{WI}, we have
$$W_{\mathcal A}\ge -k_BT I_{\mathcal A}; W_{\mathcal B}\ge\Delta F -k_BTI_{\mathcal B},$$
where $\Delta F=F_0-F_1$ is the difference in Helmholtz free energy
between the states. In the both cases, the bound for the dissipated
heat $Q_{dis}=W-\Delta F$ is
$$Q_{dis}\ge-k_BT I=-k_BT \ln P^\dagger.$$

On the other hand, the Landauer's principle gives us the inequality
$$\langle Q_{dis}\rangle=q_0Q_{dis,{\mathcal A}}+q_1Q_{dis,{\mathcal B}}\ge k_BTI_S$$
for a ``mixed'' process $q_0{\mathcal A}+q_1{\mathcal B}$. Here
$q_0$ and $q_1$ are (arbitrary) probabilities, such that
$q_0+q_1=1$, and $I_S=-q_0\ln q_0-q_1\ln q_1$ is the erased Shannon
information. This inequality is supposed to be true for \emph{any}
probabilities $q_0$ and $q_1$, which is equivalent to the pair of
inequalities
$$Q_{dis,{\mathcal A}}\ge -k_BT\ln p_0,\,Q_{dis,{\mathcal B}}\ge -k_BT\ln p_1$$
for \emph{some} positive numbers $p_0$ and $p_1$ satisfying
$p_0+p_1=1$. So, the Landauer's principle gives us exactly the same
information, except for it does not tell us where the mysterious
numbers $p_0$ and $p_1$ come from. In fact, $p_0=P({\mathcal
A}^\dagger),\,p_1=P({\mathcal B}^\dagger)$.

\begin{figure}
  \centering
  \includegraphics{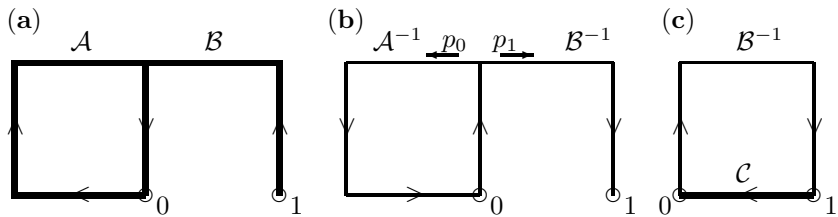}\\
  \caption{Erasure of information.}
\end{figure}

If the process ${\mathcal B}$ is reversible,  then $W_{\mathcal
B}=\Delta F -k_BT\ln p_1$. But this process is not reversible in
standard thermodynamics. We may as well consider a reversible in
standard thermodynamics process ${\mathcal C}$  with the initial
state ``one'' and the final state ``zero'', but it is a different
process which is  \emph{not} erasure.  (It means that after
${\mathcal C}$ is complete,  it is still possible to find out that
the initial state was ``one''.) In terms of nonstandard
thermodynamics, ${\mathcal C}$ is a  bideterministic  reversible
process; the work done on the system is $W_{\mathcal C}=\Delta F$.
The cycle ${\mathcal C}{\mathcal B}^{-1}$ deserves some attention
(Fig. 6 (c)). It is a process in which heat is ``antidissipated'',
$W_{{\mathcal C}{\mathcal B}^{-1}}=k_B\ln p_1<0$ . Several authors
have pointed out that this cycle is basically equivalent to the
Szilard engine cycle \cite{IF}\cite{Par}. In a sense, this is the
opposite to erasure of information: we have $I>0$ instead of $I<0$.
The original Landauer's principle does not handle this case, but it
is possible to  consider it within a more general framework proposed
by Maroney \cite{Mar}.

\subsection*{F. The magnetization reversal process}

A small magnet is one of the standard implementations of a memory
device. In view of the Landauer's principle it was considered by
Landauer himself \cite{Lan}, by Bennett \cite{B}, by  Parrondo
\cite{Par} and, more recently, by Lambson et. al \cite{LCB}. Such a
system can be described by the Hamiltonian
$$H(x)-HM(x),$$
where $H$ is external magnetic field, $x$ is a point in the phase
space and $M(x)$ is the (stochastic) magnetic moment. ($H$ is in
units of magnetic induction, i.e. $\mu_0=1$.)

We do not need any specific information about this Hamiltonian,
besides some properties. There is a symmetry $x\leftrightarrow x^*$
of the phase space (time reversal), such that $H(x^*)=H(x)$ and
$M(x^*)=-M(x)$. At a fixed temperature $T$, the moment $M=\langle
M(x)\rangle$ is a function of $H$. We assume that there is
spontaneous magnetization. It means that in the region $|H|< H_c$
there are two (symmetric) phases with different magnetic moments, so
we have a hysteresis loop on the $M$ vs. $H$ diagram (Fig. 7). (We
assume a single domain; otherwise the loop might look quite
different.)

 Consider a quasistatic magnetization reversal process ${\mathcal
A}$, which begins at $H=H_0$ and ends at $H=-H_0$ (such that
$H_0>H_c$).  In standard thermodynamics, it is an irreversible
process. At $H=-H_c$ the system undergoes an irreversible first
order phase transition, and its entropy increases by $\Delta_+S$.
Due to the symmetry, the entropy at $H=\pm H_0$ is the same, hence
$\Delta_+S+Q\slash T=0,$ where $Q$ is heat taken in the process.

The symmetric process ${\mathcal B}$ begins at $H=-H_0$ and ends at
$H=H_0$. Work done on the system in the cycle ${\mathcal A}{\mathcal
B}$ is equal to the area $A$ of the hysteresis loop, because $\dbar
W=-MdH$. Heat taken in ${\mathcal A}$ is a half of heat taken in the
cycle, i.e.  $Q=-A\slash 2$, hence
$$\Delta_+S=A\slash 2 T.$$

Nonstandard thermodynamics has a somewhat deferent view on the
process. It is postulated that there exists the opposite process
${\mathcal A}^\dagger$. Its probability $p=P({\mathcal A}^\dagger)$
can be very small, but it cannot be zero. The opposite precess is
peculiar because at $H=-H_c$ the system undergoes a backward phase
transition, and its entropy \emph{decreases}. (Which may be
interpreted as a sort of fluctuation.) Note that when $H$ increases,
the system undergoes either ${\mathcal A}^\dagger$ or ${\mathcal
B}$, hence $P({\mathcal B})=1-p$.

But it means that, by symmetry, ${\mathcal A}$ is not a
deterministic process either: $P({\mathcal A})=1-p$. At $H=H_c$ the
system can undergo the same backward phase transition and follow the
process ${\mathcal B}^\dagger$ instead of ${\mathcal A}$ (Fig.7
(b)).

\begin{figure}
  \centering
  \includegraphics{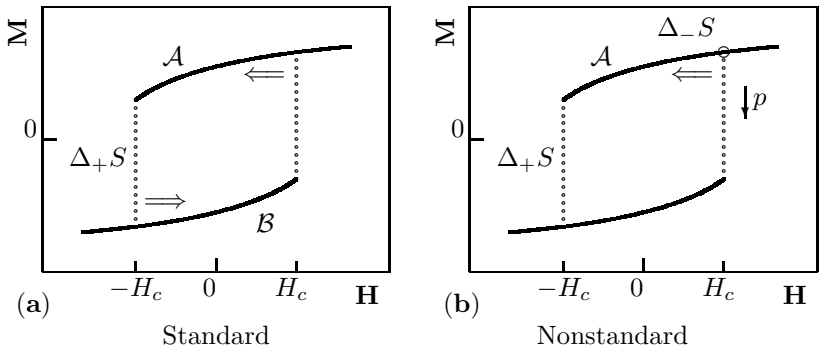}\\
  \caption{Thermodynamics of magnetization reversal process.}
\end{figure}

One can see that in the magnetization reversal process ${\mathcal
A}$ there are two points where $I\neq 0$. At $H=H_c$, we have the
forward probability $1-p$ and the backward probability one, hence
$I=-\ln(1-p)$. At $H=-H_c$ the forward probability is one, and the
backward probability is $p$, hence $I=\ln p$. By \eqref{SI}, the
entropy changes at these points are respectively
$$\Delta_-S=k_B\ln(1-p),\,\Delta_+S=-k_B\ln p.$$
On the other hand, $\Delta_+S+\Delta_-S+Q\slash T=0$, and
$Q=-A\slash 2$. Thus,

\begin{equation}
A=2k_BT\ln(p^{-1}-1). \label{hys}
\end{equation}

Note that this magnetization reversal process, which is irreversible
in standard thermodynamics, is in fact reversible in nonstandard
thermodynamics. The apparent irreversibility is a consequence of
hysteresis. A similar idea was suggested in statistical physics by
Jarzynski \cite[Sec. 4]{JJ}.

Parrondo \cite{Par}, using the analogy between a magnetic memory and
the Szilard engine, concluded that the area of the hysteresis loop
is bounded from below by $A\ge 4k_BT\ln 2$. This is a very
interesting mistake. If $A<4k_BT\ln 2$ then $p>1\slash 5$. It
follows that the process $A$, introduced in \cite[Sec. IV]{Par}, is
far from deterministic, which was neglected in \cite{Par}. Once it
is taken into account, the term $k_BT\ln 2$ on the right hand side
of Eq.(8) turns into $k_BT\ln(2-2p)$. In the end, we only have $A\ge
4k_BT\ln(2-2p)$, which is a trivial consequence of \eqref{hys}.

\section*{IV. PRINCIPLES OF NONSTANDARD THERMODYNAMICS}

\subsection*{A. Heat and work}

Interpretation of heat in nonstandard thermodynamics is a delicate
matter. Consider the definition of Clausius entropy

\begin{equation}
\Delta S=\int\frac{\dbar Q}{T}-k_BI. \label{SI2}
\end{equation}

To find the entropy difference between two states, we have to
measure the right hand side in some  reversible process. The
question is, what exactly is heat in \eqref{SI2}? The standard
interpretation of heat is simply energy taken from a heat bath. But
in fact, energy goes constantly back and forth between a system and
a bath due to heat fluctuations. Usually, the contribution of heat
fluctuations to the reduced heat  term in \eqref{SI2} is compatible
to the information term, if not much bigger. So, it is not at all
clear how to find entropy with high enough precision, so it would
make sense to take the information term into account. In statistical
physics it is not a problem, because the meaning of ``heat'' (or
``work'' etc.) is mostly ``the average of heat''. Of course, to take
average is an easy way to get rid of heat fluctuation, but average
is alien to classical thermodynamics.

There is another way: to measure \emph{work} instead of \emph{heat}.
It makes a difference, because there are no work fluctuations in a
reversible process. (Apparently, this fact is known to experts, but
the author can't produce a single reference where it is pointed out
explicitly.) The thermodynamic meaning of this proposal is to use
Helmholtz free energy $F$ instead of entropy as a basic
thermodynamic function. Other thermodynamic functions are derived
from $F$, for example
$$S=-\partial F\slash\partial T,\,U=F+TS,$$
etc. Then, heat should be defined by $Q=\Delta U-W$. Call this kind
of heat \emph{thermodynamic}, to distinguish it from
\emph{stochastic} heat $Q^*$, which is just energy taken from a heat
bath or baths. Stochastic heat is $Q^*=\Delta E-W$, where $E$ is the
energy of a system. The difference between $U$ and $E$ is that the
former is a function of a macroscopic state only while the latter is
varying due to heat fluctuations.

In this scheme, \eqref{SI2} is not a definition of entropy, it is
basically a definition of Helmholtz free energy, which can be
rewritten as

\begin{equation}
dF-\frac{\partial F}{\partial T}dT=\dbar W+k_BT\dbar I. \label{df}
\end{equation}

In this equation, no average is necessary. Unfortunately, $F$ is not
defined by \eqref{df} completely: the left hand side does not change
if it is replaced by $F+f(T)$, where $f$ is an arbitrary function.
(One cannot find specific heat by measuring work.) The only way to
resolve this uncertainty is to measure the internal energy directly.
To do this properly we have to detach the system from a heat bath
and to measure its energy $E$ in an adiabatic process; then
$U=\langle E\rangle$. At this point, one cannot avoid the average,
but it is enough to measure specific heat in a \emph{single} state,
when the only varying parameter is temperature.

Now we can give a more accurate thermodynamic interpretation of the
Jarzynski equality \eqref{J}, then in III D. Consider a quasistatic
process, reversible or not. In such a process, the distribution
remains canonical (possibly not ergodic), hence $s=(E-F)\slash T$.
By definition,
$$\dbar\sigma=ds-\frac{\dbar Q^*}{T}+k_B\dbar I,$$
where $\sigma$ is stochastic entropy production and $Q^*$ is
stochastic heat. (Which was denoted in III D by $Q$, but this must
not led to a confusion.) The equivalent equation is
$$dF-\frac{\partial
F}{\partial T}dT-k_BT\dbar I=\dbar
W-T\dbar\sigma+\left(s-\frac{\partial F}{\partial T}\right)dT.$$
Note that in this equation the terms on the left hand side are
thermodynamic while the terms on the right hand side are stochastic.
In a quasistatic process, temperature is supposed to change
gradually, hence the term $(s-\partial F\slash\partial T)dT$ can be
replaced by its average value, which is zero

\begin{equation}
dF-\frac{\partial F}{\partial T}dT-k_BT\dbar I=\dbar W-T\dbar\sigma.
\label{df2}
\end{equation}

Denote by $\Sigma=\langle\sigma\rangle$ the \emph{macroscopic}
entropy production. From \eqref{df2}, we have

\begin{equation}
\Delta S=\left\langle \int\frac{\dbar Q}{T}\right\rangle-k_B
I+\Sigma, \label{claI3}
\end{equation}

where  $\dbar Q$ is \emph{thermodynamic}  heat. For a cyclic
process, the consequence of \eqref{claI3} is an analog of
\eqref{claI2}, because $\Sigma\ge 0$. The important difference is,
in \eqref{claI3} the average is over work fluctuations while in
\eqref{claI2} the average is over heat fluctuations. If a process is
reversible, then $\Sigma=0$ and \eqref{claI3} turns into
\eqref{SI2}; in this case we do not need any average at all.

Another important consequence of \eqref{df2} is that the variables
on the right hand side must have the same dispersion $D(\dbar
W)=T^2D(\dbar\sigma)$. By assumption, the process is quasistatic,
hence the contributions to $\sigma$ from different parts of this
process must be statistically independent. So, $\sigma$ must have a
normal distribution. From the Jarzynski equality \eqref{J}, we have
$D\sigma=2k_B\langle\sigma\rangle$. Moreover, this argument is valid
not only for the whole process, but also for any part of it, hence
$D(\dbar\sigma)=2k_B\dbar\Sigma$. Taking into account that
dispersion is additive, we have

\begin{equation}
DW=2k_B\int T^2\dbar\Sigma.\label{DW}
\end{equation}

For example, if the process is reversible, then $DW=0$. (Presumably,
a truly quasistatic process is reversible for a sensible model. But
the equalities \eqref{claI3} and \eqref{DW} must be valid for an
irreversible process which is close to quasistatic.)

\subsection*{B. The Szilard engine}

The Szilard engine is an imaginary device invented by L. Szilard in
1929 \cite{S}.  The thermodynamics of the engine was discussed in
the literature: many times in terms of statistical physics
\cite{B,SU,Fah,Par,MNV} and one time in terms of classical
thermodynamics (essentially)\cite{IF}.

The engine consists of a box with a single particle. The box is
provided with a thin piston which can be inserted to or removed from
it as necessary (Fig 8). The engine undergoes an isothermal cyclic
process in contact with a heat bath at temperature $T$. At the
beginning, the piston is out of the box. As the first step of the
process it is inserted into the box at the middle, dividing it into
two parts of equal volume. The particle gets trapped in one of the
halves. After that, the piston moves into the empty half until it
reaches the wall. The piston is then removed and the cycle is
complete.

\begin{figure}
  \centering
  \includegraphics{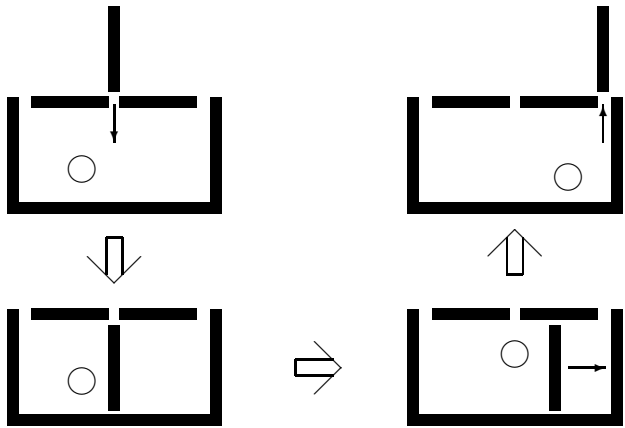}\\
  \caption{The Szilard engine}
\end{figure}

Under natural assumptions $P=k_BT/V$ by the Gay-Lussac law, and the
work done in the cycle on the system is negative
$$W=-\int PdV=-k_BT\ln 2,$$
in apparent contradiction to the second law. The correct solution of
the paradox, found by Bennett \cite{B}, is well known, so there is
no need to discuss it here in detail. Still, there are several
important points apparently missed in the literature.

To begin with, there are two different ``paradoxes'' to deal with: a
violation of the standard Clausius inequality (which is real) and a
violation of the second law (only apparent). Admittedly, a violation
of \eqref{cla} is not really a problem. The cycle is not
deterministic; we simply do not have a reason to expect the
inequality to be true in such an exotic situation. Also, the
explanation in nonstandard thermodynamics is straightforward. There
are two different cyclic processes, each with probability $P=1\slash
2$ and the backward probability $P^\dagger=1$, hence thermodynamic
information created in any of the cycles is $I=\ln 2$. By
\eqref{claI}, we have
$$Q=k_BT\ln 2,$$
in perfect agreement with the Gay-Lussac law. (Note that the
processes are reversible.) One may also consider a ``skewed''
version of the engine, where the partition is not inserted in the
middle  \cite{MNV}. If the particle occupies volume $V_0$ and the
volume of the empty half is $V_1$, then $P=V_0\slash(V_0+V_1)$ and
$$Q=k_BT\ln(1+V_1\slash V_0),$$
again in agreement with the Gay-Lussac law.

The apparent contradiction to the second law is a more delicate
matter. In short, the Bennett's argument is as follows. To work
properly, the engine needs a controller and a controller needs a
memory. After a cycle, 1 bit of information has to be erased from
this memory, and all the heat returns to the bath by the Landauer's
principle. This is very similar to the environment dependence
argument considered in II A: a memory plays the role of a
``parallel'' system $S^\prime$. (We have $Q=k_BT\ln 2$,
$Q^\prime=-k_BT\ln 2$, and $Q+Q^\prime=0$.)

However, it is not exactly the same argument. In fact, the both
Szilard engine cycles are environment independent. They only become
``look'' environment dependent when considered together, as a single
mixed process. (The engine does not need any controller to work
properly in \emph{one} of the cases!)  The Bennett's explanation is
completely sound, but it cannot be accepted as a proper explanation
in nonstandard thermodynamics, because it is inherently
probabilistic. To make sense of it, one has to consider
\emph{ensembles} instead of a single engine undergoing a single
process. The right explanation in nonstandard thermodynamics is a
\emph{reformulation} of the second law, considered below.

Another subtle point is the behavior of the total entropy. In the
Szilard engine cycle, thermodynamic information $I=\ln 2$ is created
at the moment when the partition is inserted. By \eqref{SI}, the
entropy of the engine decreases instantly by $k_B\ln 2$, which is
what one might expect when the volume is halved. The problem is, the
entropy of any other system in the environment remains the same,
which means that the \emph{total} entropy decreases as well. This
paradox was discussed in \cite{Par} and in \cite{IF}.  Parrondo
proposed a redefinition of the entropy to solve the problem. This
proposal (in different interpretation) is discussed in IV D.

\subsection*{C. The second law revised}

It has already been mentioned that determinism is among the basic
assumptions in standard classical thermodynamics. However, it does
not mean that indeterministic phenomena cannot be considered at all:
to some extent, they can. Apparently, the \emph{de facto}
interpretation of the second law in this situation is what one might
call a statistical formulation of the law. It has never been
formulated explicitly, for the best of author's knowledge. We
propose the following formulation:  \emph{the average of heat taken
from a single heat bath and converted to work in a mixed global
cyclic processes is not positive}. It means, if heat $Q_i$ is taken
in a process ${\mathcal A}_i$, which is observer with probability
$P_i$, then
$$\sum_i P_i Q_i\le 0$$

(here $P_i=P({\mathcal A}_i)$ and $\sum_i P_i=1$.)

This formulation of the second law is not very satisfactory. To make
any conclusion about a particular process which is not deterministic
($P<1$), we have to take into account every possible alternative
process. Moreover, for the same process the list of alternatives can
be different in different situations. A more reasonable formulation
of the law is proposed below.

Consider a global cyclic process ${\mathcal A}$ with forward
probability $P$ and backward probability $P^\dagger$, in which heat
$Q$ is taken from a single heat bath at temperature $T$ and
converted to work. It is formally possible to treat all the systems,
except for this bath, as a \emph{single} big system. The inequality
\eqref{claI} is valid for this big system, which means

\begin{equation}
\frac{Q}{T}\le k_B\ln\frac{P^\dagger}{P}. \label{QTI}
\end{equation}

In the case $P\ge P^\dagger$ the right hand side is not positive.
Thus, \emph{it is not possible to take heat from a single heat bath
and convert it to work in a balanced global cyclic process}. Call
this statement \emph{a general Kelvin-Planck formulation} of the
second law. It is, in fact, very close to the original (special)
formulation: it is enough to replace  \emph{balanced} by
\emph{deterministic}.

This formulation of the second law was obtained as a consequence of
the fluctuation theorem. However, once it is taken for granted, we
need no more appellations to statistical mechanics. Consider again
the global process ${\mathcal A}$. It $P<P\dagger$, then we cannot
apply the (new formulation of) the second law directly. But we can
introduce another process  ${\mathcal C}$, which is the reverse to
the ``skewed'' Szilard engine cycle described in IV B, such that
$V_0\slash(V_0+V_1)=P\slash P^\dagger$. We have $P({\mathcal
A}{\mathcal C})=P(({\mathcal A}{\mathcal C})^\dagger)$, hence the
cycle ${\mathcal A}{\mathcal C}$ is balanced and $Q+Q_{\mathcal
C}\le 0$. Heat  $Q_{\mathcal C}$, taken in the process ${\mathcal
C}$, can easily be found from the Gay-Lussac law
$$Q_{\mathcal C}=k_BT\ln(V_0\slash (V_0+V_1))=-k_BT\ln\frac{P^\dagger}{P},$$
and the inequality \eqref{QTI} follows. To prove the same inequality
in the case $P>P\dagger$, one can use the process ${\mathcal
C}^{-1}$.

Assuming, as usual, environment independence, we can deduce
\eqref{claI} from \eqref{QTI}, following Clausius. (Note that the
logic is reversed: the general case is a consequence of a special
case.) So, we can use the general formulation of the second law as a
fundament of nonstandard thermodynamics, basically repeating the
standard scheme. In this argument the Szilard engine plays
essentially the same role of a perfect machine as the Carnot engine
in traditional thermodynamics. The statistical formulation of the
second law should not be considered a fundamental principle, it is a
consequence of the general Kelvin-Planck formulation of the law.
From \eqref{QTI}, we have the  inequality
$$\left\langle e^{Q\slash k_BT}\right\rangle\le \sum_i P_i^\dagger=1,$$
which resembles the Jarzynski equality. By the Jensen inequality,
$\langle Q\rangle\le 0$.

Returning to the question considered in IV B, the Szilard engine
cycle does not contradict any of the formulations of the second law.
It does not contradict the standard Kelvin-Planck formulation,
simply because it cannot be applied to a process which is not
deterministic. It does not contradict the statistical formulation by
the Bennett's argument. And it obviously does not contradict the
general Kelvin-Planck formulation.

\subsection*{D. The total entropy}

It is often convenient \emph{not} to call a global process cyclic
unless $P=P^\dagger$. Apparently, this trick does not make much
sense in general, but when it is possible, it allows to introduce a
new thermodynamic function $H$, which can be interpreted as
\emph{stored} thermodynamic information.

The probability of a global process is equal to the product of the
probabilities of the local ones. It follows that the condition
$P=P^\dagger$ is a restatement of
$$\oint\sum_i\dbar I_i=0,$$
where the sum is taken over all the systems.  We can pretend that
the sum is an exact differential and introduce a function $H$ by
$$dH=\sum_i\dbar I_i.$$
Following Parrondo \cite[eq. (16)]{Par}, the total entropy can then
be redefined by
\begin{equation}
S_{tot}=\sum_i S_i+k_B H, \label{Stot}
\end{equation}
where the sum is taken over all the systems, \emph{including heat
baths}.

We have the inequality \eqref{SI1} for a ``genuine'' system and the
equality $dS=\dbar Q\slash T$  for a heat bath. Thus,
$dS_{tot}\ge\sum_i\dbar Q_i\slash T_i$. But the sum on the right
hand side is equal to zero. Heat taken \emph{by} a system is heat
taken \emph{from} a bath, so all the terms in this sum are canceled.

So, $S_{tot}$ defined by \eqref{Stot} is a nondecreasing function,
unlike the ``naive'' total entropy which does not include the term
$k_B H$. This term can formally be interpreted as the  entropy of a
fictitious system, called a \emph{buffer}. This is convenient,
because we can use the language of weak thermodynamics; that is, we
can pretend that the process is deterministic. Thermodynamic
information must then be interpreted as entropy, transferred
adiabatically from a system to the buffer
$$\dbar{\mathfrak S}=-k_B\dbar I,$$

and the general Clausius inequality \eqref{claI} must be considered
a special case of \eqref{cla2}.

\subsection*{E. Thermodynamics of the Szilard engine}

Taking into account the entropy of the buffer allows us to take a
fresh look at nonstandard thermodynamics. We consider below the
Szilard engine as an example. The author would like to point out
some essential differences from what one may find in the literature.
We consider what happens if the particle gets into a particular part
of the engine. What happens otherwise is ignored, as it is a
different process. We follow \cite{IF}, so the main subject is the
Clausius entropy of the engine, but we give a much more detailed
account of it.

The engine, which is connected to a heat bath at temperature $T_1$,
is operated by a controller (called a Maxwell's demon). There is no
need to treat a controller as a thermodynamic system, it may well be
considered a kind of external agent instead. But a controller needs
a \emph{memory}, which certainly is a thermodynamic system. In the
cycle, it undergoes an erasure of information process, accompanied
by heat generation. The heat must be absorbed by a heat bath,
connected to the memory. This is a  \emph{different} bath, at
temperature $T_2$ (Fig. 9).

The memory is a usual 1-bit memory device considered in III E, which
is initially in ``zero'' state. After the partition is inserted, the
controller measures the position of the particle and brings the
memory to ``zero'' or ``one'', depending on where the particle
happens to be. We consider the case when the memory remains in
``zero'' state all the time (but another case is not very
different).

We do not assume that the engine or the memory are ``straight''; it
is possible that $V_0\neq V_1$ and $p_0\neq p_1$. However, we have
to assume that $p_0\slash p_1=V_0\slash V_1$. Otherwise, we would
have $P\neq P^\dagger$ for the global process, so it would not be a
proper cycle.  The natural entropy unit in this case is $u=-k_B\ln
p_0$ (which is $k_B\ln 2$ if $p_0=p_1$).

\begin{figure}
  \centering
  \includegraphics{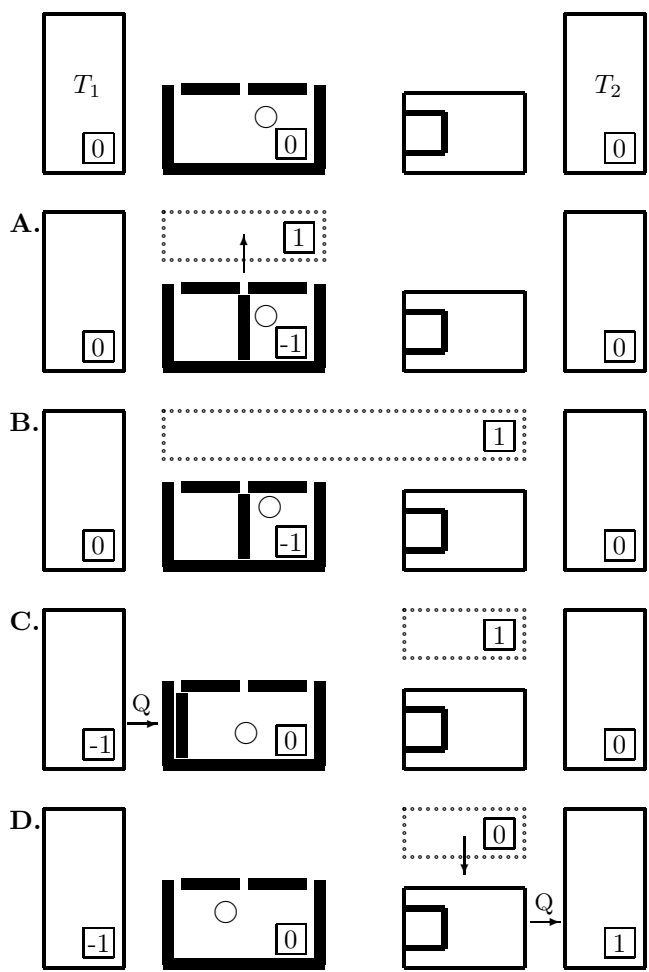}\\
  \caption{The Szilard engine cycle. The engine is connected to a
  heat bath at temperature $T_1$. The memory on the right
is connected to a  bath at temperature $T_2$. The dotted rectangle
represents the buffer.   Entropy of each system at each step is
shown in a square box. Entropy transfer is shown by arrows.}
\end{figure}

The cycle consists of four steps.
\newline
{\bf A.~ Insertion} The piston is inserted into the box, such that
the particle is trapped in the volume $V_0$.  The entropy of the
engine decreases by $u$. By now, $H=-\ln p_0$ (thermodynamic
information is stored in the engine), hence the entropy of the
buffer increases by $u$. The formal explanation is adiabatic entropy
transfer to the buffer (Fig.9).
\newline
{\bf B.~ Measurement} Actually, nothing is changed. However, it is
now more natural to assume that information is stored not in the
engine alone, put in the \emph{pair} engine $+$ memory. (Because the
states of two systems are correlated. Strictly speaking, this is not
a meaningful fact when we consider a single process.)
\newline
{\bf C.~ Expansion} The piston moves to the wall. The
single-particle gas performs work at the expense of heat taken from
the bath. The entropy of the engine increases by $u$ due to this
heat. (Now information is stored in the memory only.)
\newline
{\bf D.~ Erasure} This is the process ${\mathcal A}$ considered in
III E. In terms of ``buffer thermodynamics'', entropy is transferred
from the buffer to the memory. The entropy of the memory does not
actually increase because of heat exchange with the bath.

The net result of the whole cycle is adiabatic entropy transfer
between two heat baths, mediated by the buffer. Work done on the
engine is $-u T_1$, work done on the memory is $u T_2$. Basically,
what we have is a heat engine, but heat undergoes an unusual
transformation
$${\rm heat}\Longrightarrow{\rm work}+{\rm thermodynamic \,\,\,information}\Longrightarrow
{\rm heat}.$$

\section*{V. CONCLUSIONS}

In this paper the foundations of the Clausius inequality are
discussed. It is shown that, strictly speaking, the inequality is
not a consequence of first principles of thermodynamics. So, the
validity of the inequality in general should be considered an open
problem, not an established fact.

To put it into other words, if we do not make as many implicit
assumptions as usual, then a hypothetical violation of the Clausius
inequality does not contradict any of the known laws of physics. It
can be explained by adiabatic entropy transfer between two systems
(not accompanied by energy transfer or matter exchange). On this
basis, one can build a consistent version of classical
thermodynamics, called \emph{weak thermodynamics}.

Whether weak thermodynamics has something to do with the physical
reality or not, it may at least be useful for some purposes. The
Clausius inequality is a falsifiable statement. One cant test it in
a laboratory; in this respect, weak thermodynamics may play a
similar role to the post-Newtonian formalism which is a tool in
tests of general relativity. Also, it may be of help in analysis of
exotic processes, which could not be analyzed properly within the
standard thermodynamic framework.

In this paper, an extension of classical thermodynamics is proposed.
This theory, called \emph{nonstandard thermodynamics}, is supposed
to describe thermodynamic processes which may be observed with some
probability. This includes ergodicity breaking, some kind of
fluctuations, and other phenomena beyond the scope of standard
thermodynamics. The principal tool is a generalized Clausius
inequality \eqref{claI}, which may be considered one of the possible
formulations of the Landauer's principle. To a large extent,
nonstandard thermodynamics is an attempt to translate some well
known results, like fluctuation theorems, from the language of
statistical physics to the language of classical thermodynamics. It
turns out that weak thermodynamics and nonstandard thermodynamics
have much in common. In fact, it is sometimes convenient to take
them for the same theory.

The generalized Clausius inequality \eqref{claI} is a direct
consequence of the generalized Jarzynski equality
\cite[Eq.(7)]{KPV}. This equality has a solid theoretical basis and
was tested in feedback control experiments \cite{T}. For this
reason, there is little doubt that the standard Clausius inequality
\eqref{cla} can be violated in a feedback control experiment,
although it may be difficult to prove this directly. If this
inequality can be violated on macroscale is a more difficult
question, which deserves some attention.

\end{document}